\documentstyle[12pt]{article}
\newlength{\abstractwidth}
\begin{document}
\thispagestyle{empty}
\pagestyle{plain}
\def\half{{\textstyle{1\over2}}}
\def\sq{{\textstyle{{{1}\over{\sqrt 2}}}}}
\def\sqtwo{{\textstyle{{{1}\over{2 \sqrt 2}}}}}
\def\squ{{\textstyle{\sqrt 2}}}
\renewcommand{\thefootnote}{\fnsymbol{footnote}}
\renewcommand{\thanks}[1]{\footnote{#1}} 
\newcommand{\starttext}{
\setcounter{footnote}{0}
\renewcommand{\thefootnote}{\arabic{footnote}}}

\begin{titlepage}
\bigskip
\hskip 3.7in\vbox{\baselineskip12pt
\hbox{NSF-ITP-95-50}\hbox{hep-th/9506048}}
\bigskip\bigskip\bigskip\bigskip

\centerline{\large \bf Moduli Space of CHL Strings}

\bigskip\bigskip
\bigskip\bigskip

\centerline{\bf Shyamoli Chaudhuri\thanks{sc@itp.ucsb.edu}
and
Joseph Polchinski\thanks{joep@itp.ucsb.edu}}
\medskip
\centerline{Institute for Theoretical Physics}
\centerline{University of California}
\centerline{Santa Barbara, CA\ \ 93106-4030}

\bigskip\bigskip

\begin{abstract}
\baselineskip=16pt
We discuss an orbifold of the toroidally compactified heterotic
string which gives a global reduction of the dimension of the
moduli space while preserving the supersymmetry. This construction
yields the moduli space of the first of a series of reduced rank
theories with maximal supersymmetry discovered recently by
Chaudhuri, Hockney, and Lykken. Such moduli spaces contain
non-simply-laced enhanced symmetry points in any spacetime
dimension $D$$<$$10$. Precisely in $D$$=$$4$ the set of allowed
gauge groups is invariant under electric-magnetic duality,
providing further evidence for S-duality of the $D$$=$$4$ heterotic
string.
\end{abstract}
\end{titlepage}
\starttext
\baselineskip=18pt
\setcounter{footnote}{0}

\section{Introduction}

Toroidal compactification of the heterotic string~\cite{narain}
preserves the full spacetime supersymmetry, giving the algebras
$N$$=$$4$ in $D$$=$$4$, $N$$=$$2$ (nonchiral) in $D$$=$$6$, and
$N$$=$$1$ in $D$$=$$8$. These maximally supersymmetric
compactifications have played a major role in recent discussions of
strongly coupled string theory.\footnote {For reviews and recent
discussions see refs.~\cite{sdual},
\cite{ssdual},~\cite{eddual}.}

Banks and Dixon\cite{bd} showed that maximal supersymmetry in the
heterotic string requires that the right-moving (supersymmetric)
target space be a
$(10$$-$$D)$-dimensional torus.  For some time the only known
example was that of ref.~\cite{narain} in which the left-moving
degrees of freedom were also toroidal.  Recently, Chaudhuri,
Hockney, and Lykken (CHL) have used fermionization to construct
many new exact conformal field theory solutions\footnote{We will
use the term \lq\lq solutions'' since it not clear at present
whether all solutions to the heterotic string consistency
conditions can be interpreted as backgrounds or compactifications
of the ten-dimensional string, though this is widely assumed.}
having maximal supersymmetry in any spacetime dimension
$D$$<$$10$~\cite{chl}. These solutions are characterized by a
reduction of the rank of the left-moving gauge group relative to
the toroidally compactified ten dimensional heterotic string.  CHL
have found maximally supersymmetric $N$$=$$4$ theories in $D$$=$$4$
spacetime dimensions with gauge groups of rank
\begin{equation}
(r_L, r_R)=(26-D-k,\ 10-D),  \quad k=8,12,14,16,18,20,22 ~ .
\end{equation}
Here $k$ is the reduction of the rank. The first two members of
this series can be followed up to (even) spacetime dimension
$D$$=$$8$ and $D$$=$$6$, respectively.  Since the moduli in these
maximally supersymmetric theories appear in vector multiplets, the
number of inequivalent marginal deformations at any point in the
moduli space is fixed to be $r_L r_R$.  It was shown in
\cite{chl},\cite{sc} that such moduli spaces contain points of both
non-simply-laced and simply-laced enhanced symmetry, as well as
higher level realizations of the gauge symmetry.\footnote
{We should also note that analogous solutions for the Type II
string, four dimensional theories with extended spacetime
supersymmetry including
$N$$=$$4_R$$+$$ 0_L$, $4_R$$+$$ 1_L$,
$4_R $$+$$ 2_L$, and
$2_R $$+$$ 2_L$, and with reduced dimension moduli spaces, were
discovered some time ago by Ferrara and Kounnas using the free
fermionic construction~\cite{costas}. The Type II solutions differ
from those of
\cite{chl} in using left-right symmetric Majorana fermions, which
are excluded in the heterotic case by maximal supersymmetry
\cite{bd}. The heterotic solutions require chiral Majorana
fermions. A target space interpretation of these Type II conformal
field theory backgrounds, and the identification of their duals,
remains to be explored. We would like to thank Costas Kounnas for
informing us of this work.}

The purpose of this paper is to clarify the nature of these
theories.  While fermionization has proven to be a powerful
technique for discovering qualitatively new classes of string
compactifications it can give only isolated pieces of their moduli
space, generally at points of higher symmetry.  We will describe an
orbifold construction which yields the moduli space of the first
theory in this series, recovering, in particular, all of the $k=8$
solutions given by the fermionic construction.  We also discuss the
implications for strong-weak coupling duality in string theory.  For
the conjectured self-duality of the heterotic string in $D=4$ we
find new evidence: gauge groups always appear with their
electric-magnetic duals, for example $Sp(20)
\times SO(9)$ and $Sp(8) \times SO(21)$ in the same moduli space.
However, the CHL theories present a puzzle for the conjectured
string-string duality in $D=6$.

\section{Construction}

A clue to the nature of these theories comes from considering the
decompactification limit.  Moduli $j \tilde j$ can always be
constructed from the Cartan subalgebra  of the gauge group. At
large values of these moduli the spectrum becomes continuous, and
as long as the left-moving rank is at least
$10-D$ one can reach in this way a theory with 10 large
translationally invariant dimensions, which must be a toroidal
compactification of either the $E_8$$\times$$E_8$ or $SO(32)$
heterotic string.   This seems a puzzle at first, since the
construction of ref.~\cite{narain} appears to be general, but there
is at least one additional possibility.  Consider the effect of
transporting a string around a toroidal dimension.  The string must
return to its original state up to a symmetry $g$.  For
$g$ a general gauge transformation (inner automorphism of the gauge
group), this is equivalent to a Wilson line background and is
therefore included in the classification of ref.~\cite{narain}.

However, the $E_8 \times E_8$ theory also has an outer isomorphism
which interchanges the two $E_8$'s.\footnote {The outer isomorphism
has been considered previously in refs.~\cite{outer}, and very
recently in \cite{andy}, but always in conjunction with right-moving
twists which reduce the spacetime supersymmetry.} Modding out by
the outer isomorphism alone does not lead to a new theory, because
the gauge bosons of the other $E_8$ are recovered in the twisted
sector. The only additional possibility in $D$$=10$ is to combine
this action with a non-trivial twist on the world-sheet fermions,
but this breaks the spacetime supersymmetry to
$N$$=$$0$ \cite{outer}. Since our interest is in $Z_2$ actions that
leave the supersymmetry unbroken, twists on the world-sheet
fermions are excluded. Compactifying on a torus to any spacetime
dimension $D$$<$$10$ opens up a new option: an accompanying
translation in the torus.  Begin with an ordinary toroidal
compactification to $D$ spacetime dimensions and twist by $RT$,
where $R$ is the outer isomorphism that interchanges the two $E_8$
lattices and $T$ is a translation in the spacetime torus.  This has
just the necessary action: it eliminates one linear combination of
the two $E_8$'s, leaving the diagonal $E_8$ at level~2.

By a similarity transformation
$T$ and $R$ can be taken to commute.  Without loss of generality we
assume that $(RT)^2 = T^2$ is a symmetry of the original lattice,
else we could twist by $T^2$ to abtain a different lattice in the
same moduli space. Also we can assume that $T$ is not a symmetry of
the lattice; if it were then $RT$ would be equivalent to
$R$ and so would actually act trivially.

Denote a general momentum state by $|p_1,p_2,p_3\rangle$.
Here
\begin{equation}
p_1^I = \sq (p^I - p'^I), \quad
p_2^I = \sq (p^I + p'^I), \qquad I = 1, \ldots, 8
\end{equation}
are the linear combinations of momenta in the two $E_8$
lattices which are respectively reflected and left invariant
by $R$, while
\begin{equation}
(p_{3}^m ; \tilde p_{3}^m), \qquad m = D, \ldots, 9
\end{equation}
are the momenta of the torus.  Taking $T$ to be a translation by
$2\pi(0, v_2^I, v_{3}^m ; \tilde v_{3}^m)$, $RT$ acts as
\begin{equation}
RT |\, p_1,\ p_2,\ p_3\rangle = e^{i2\pi v \cdot p}
|-p_1,\ p_2,\ p_3\rangle
\end{equation}
where the inner product has signature~$(26-D, 10-D)$.

We start with the $E_8$$\times$$E_8$ theory,
with $10 - D$ dimensions compactified on a given torus without
background gauge fields.
The momentum lattice of such a theory is of the form
\begin{equation}
\Gamma = \Gamma^8 \oplus \Gamma^8 \oplus \Gamma',
\end{equation}
where $\Gamma'$, is even, self-dual, and of Lorentzian
signature $(10$$-$$D,10$$-$$D)$.
In the basis above $\Gamma$
takes the form
\begin{equation}
\left| \sq (p^I - p'^I),\ \sq (p^I + p'^I),\ p_3 \right\rangle,
\end{equation}
where $p^I, p'^I \in \Gamma^8$ and $p_3 \in \Gamma'$.
Specifically,
\begin{equation}
\Gamma^8: \quad p^I = \half m^I, \qquad m^I \in {\bf Z},\ \ m^I -
m^J
\in 2{\bf Z},\ \ \sum_I m^I \in 4{\bf Z}.\nonumber\\
\end{equation}

General points in the moduli space can then be reached by boosts of
the momentum lattice as in ref.~\cite{narain}. In order to allow a
twist by $R$ the boost $\Lambda$ must commute with
$R$, leaving $p_1$ invariant.\footnote {More generally there is the
possibility that $\Lambda R \Lambda^{-1}$ be a nontrivial discrete
symmetry (duality) of $\Gamma$, leading to a disconnected moduli
space.  We will not consider this here.} This subgroup is $SO(18 -
D, 10 - D)$, so the moduli space of inequivalent vacua is locally
of the form
\begin{equation}
{SO(18-D, 10-D)}\over{SO(18-D)\times SO(10-D)}
\end{equation}
as required by considerations of low energy supergravity.

The twist by $RT$ produces an asymmetric orbifold \cite{nsv}.  We
review the relevant results from that paper.  The lattice $I$ is
defined to consist of those momenta invariant under $R$.  Here, this
implies
\begin{equation}
I:\quad p_1 = 0,\ \ p_2 \in \squ \Gamma^8,\ \ p_3 \in \Gamma'.
\end{equation}
The dual lattice, in the subspace invariant under $R$, is
\begin{equation}
I^*:\quad p_2 \in  \sq \Gamma^8 ,\ \  p_3
\in \Gamma'.
\end{equation}
The number of twisted sectors is $D$ where
\begin{equation}
D^2 = \det{}'(1-R) |I/I^*| = 2^{8-8} = 1.
\end{equation}
The momenta in the twisted sector are
\begin{equation}
p \in I^* + v \label{twistmom}
\end{equation}
and the twisted sector spectrum consists of the level-matched
states. The untwisted sector simply retains the $RT$-invariant
states.  The left-moving zero point energies are the usual $-
\frac{24}{24} = -1$ in the untwisted sector and $-\frac{16}{24} +
\frac{8}{48} = -\frac{1}{2}$ in the twisted sector.

\subsection{Compactification to $D=8$}

Consider first the case $D=8$.  Let
$\Gamma'$$=$$\Gamma^{SU(2)}$$\oplus$$\Gamma^{SU(2)}$, two
copies of the free boson at the $SU(2)$ radius:
\begin{equation}
\Gamma^{SU(2)} :\quad p_{3} = \sq \left( n ;\ {\tilde n}\right) ,
\qquad
n, \tilde n \in {\bf Z},\ \ n + \tilde n \in 2{\bf Z}.
\end{equation}
We take $T$ to be a translation by half
the spacetime periodicity in the $m$$=$$9$ direction,
\begin{equation}
v_2 = 0, \quad v_3 = \left( 0,\sqtwo ;\ 0,-\sqtwo \right).
\end{equation}
Let us determine the gauge symmetry.  The untwisted sector contains
ten neutral left-moving gauge bosons\footnote {The right moving
gauge symmetry is always $U(1)^{10-D}$ in a maximally
supersymmetric compactification of the heterotic string \cite{bd}.}
\begin{equation}
\sq (\alpha^I_{-1} + \alpha'^I_{-1})
|0,0,0\rangle ,
\qquad \alpha^m_{-1} |0,0,0\rangle, \quad m=8,~9, \label{neutral}
\end{equation}
the eight antisymmetric neutral combinations being removed by the
$RT$ projection. Let $r$ denote any root of $E_8$. The charged gauge
bosons in the $E_8$$\times$$E_8$ theory which are invariant under
$RT$ are the symmetric combinations
\begin{equation}
\sq \Bigl(\,\Bigl| \sq r ,\ \sq r ,\ 0 \Bigr\rangle
+ \Bigl| - \sq r ,\ \sq r ,\ 0 \Bigr\rangle\Bigr) \label{e8l2}
\end{equation}
and the $SU(2)$ gauge bosons
\begin{equation}
p_1 = p_2 = 0,\ p_3 = \left( \pm \squ , 0 ;\ 0,0 \right).
\label{su28}
\end{equation}
In the twisted sector there are no massless states because the
right-moving component of the momentum~(\ref{twistmom}) will not
vanish. The states~(\ref{e8l2}) form the $E_8$ root lattice, but the
momentum~$p_2$ to which the neutral gauge bosons~(\ref{neutral})
couple is scaled down by a factor of $\sqrt{2}$ so the current
algebra realization of $E_8$ is at level~2.  The gauge
bosons~(\ref{su28}) are the remnant of one of the original
$SU(2)$'s, the other being removed by the $RT$ projection, so that
the gauge symmetry originates in the left-moving current algebra
$(E_8)_2 \times SU(2)_1 \times U(1)$.

General points in the moduli space are reached by $SO(10,2)$ boosts.
Generically the left-moving symmetry is broken to $U(1)^{10}$, but
there are also points of enhanced gauge symmetry.  Consider for
example the vectors
\begin{eqnarray}
e_{(1)}:&& p_1 = p_2 = 0,\ p_3 = \squ \left( 1,0;\ 0,0 \right), \nonumber\\
e_{(2)}:&& p_1 = p_2 = 0,\ p_3 = \sq \left( 0,3;\ 2,1 \right).
\end{eqnarray}
These survive the $RT$ projection in the untwisted sector
and satisfy
\begin{equation}
e_{(i)} \cdot e_{(j)} = 2 \delta_{ij}. \label{orth}
\end{equation}
Making an $SO(2,2)$ boost to a frame in which $\sq e_{(i)}$ form a
left-moving orthonormal basis, these become $SU(2)$$\times
SU(2)$$=$$ SO(4)$ weights.  The vectors
\begin{equation}
\pm \half e_{(1)} \pm' \half e_{(2)}  ,
\end{equation}
where all four independent choices of sign are included, appear in
the twisted sector~(\ref{twistmom}).  These momenta have
length-squared~1 and so with the twisted zero-point energy give
rise to massless states, combining with $\pm e_{(i)}$ to form the
root lattice of $SO(5)$.  The left-moving current algebra is
$(E_8)_2 \times SO(5)_1$.

Notice in this example that short roots of a level~1
non-simply-laced algebra and long roots of a level~2 algebra both
have $p_2 \cdot p_2 + p_3 \cdot p_3 = 1$, and that they can arise
in the Hilbert space in two ways: as untwisted states with momentum
$p_1 \cdot p_1 = 1$ or as twisted states.  Note also that in this
moduli space non-simply-laced algebras can only appear with root
lengths in the ratio $1:\squ$, and only as level~1 realizations,
and that simply-laced algebras can appear only at levels~1 or~2.

To find further enhanced symmetry points it is useful to focus first
on the long roots, which have $p_1 = 0,\ p_2 \in \squ \Gamma^8$. We
will now construct an orthonormal set of long roots, with inner
product $e_{(i)}$$\cdot$$e_{(j)}$$=$$2 \delta_{ij}$, that extends
the basis $( e_{(1)}, e_{(2)})$ described above to the root-lattice
of $(SU(2))^{10}$. To begin with, we identify a useful basis of
vectors contained in the euclidean lattice $\squ \Gamma^8$. These
are
\begin{eqnarray}
u_{(i)}:&& p^1_2 = \squ , ~~~ p^{i+1}_2 = \squ , \quad i= 1, \ldots,
 7 \nonumber\\
u_{(8)}:&& p^I_2 = \sq ,  \quad I=1, \ldots 8
\end{eqnarray}
with the property $u_{(i)} \cdot u_{(j)} = 2 + 2 \delta_{ij}$.
The vectors $e_{(1)}$, $e_{(2)}$, and
\begin{equation}
e_{(i)}:\qquad p_2 = u_{(i-2)},\ p_3= (0,\squ ;\ \squ ,\squ ),
\quad i= 3,
\ldots, 10 ,
\end{equation}
satisfy the orthonormality condition~(\ref{orth}) and so by a
Lorentz transformation can be taken to a left-moving basis, at
which point they form the $(SU(2))^{10}$ root lattice normalized to
level~1. The lattice
$(SU(2))^n$ is the long root lattice of the non-simply laced group
$Sp(2n)$. The linear combinations
\begin{equation}
\pm \half e_{(i)} \pm' \half e_{(j)}
\end{equation}
are vectors of length $1$ filling out the short root lattice of the
group $Sp(20)_1$. These states can all be found in the untwisted
sector, except for $i=2$ or $j = 2$ which are contributed by the
twisted sector.

Since we lack a more elegant characterization of the allowed
momentum lattices, we will follow the above procedure in looking for
points of enhanced symmetry. Fortunately,  it suffices for the
examples at hand. The possible enhanced symmetry points that can
appear in the D=8 moduli space are limited. In the appendix we show
that at any point in this moduli space the long roots are always
orthogonal, so they can only form products of $SU(2)$'s.  The only
non-simply-laced groups that can appear are therefore $Sp(2n)$,
including $Sp(4) = SO(5)$, while the only simply laced group that
can appear at level~1 is a product of $SU(2)$'s.

Including an inner automorphism (Wilson line) in the translation
leads to nothing new.  That is, twist by $RT'$ where now
$v_2$ is nonzero.  The requirement that $T'^2$ be a symmetry of
$\Gamma$ implies that $v_2 \in I^*$, but then the asymmetric
orbifold generated by $RT'$ has the same spectrum as that generated
by $RT$.

It is interesting to consider the decompactification limit of the
$Sp(20)$ theory in which the radius of the 9-direction is taken to
infinity. Initially the gauge symmetry is broken to $Sp(18)$, but
in the limit the twist becomes irrelevant and the antisymmetric
combinations
\begin{equation}
\sq (\alpha^I_{-1} - \alpha'^I_{-1}) |0,0,0\rangle ,
\qquad
 \sq \Bigl(\,\Bigl| \sq r ,\ \sq r ,\ 0 \Bigr\rangle
- \Bigl| - \sq r ,\ \sq r ,\ 0 \Bigr\rangle\Bigr)
\end{equation}
become massless.  These lie in the antisymmetric tensor representation
of $Sp(18)$, combining with the $Sp(18)$ adjoint to form the adjoint of
$SU(18)$, a Narain compactification.  Decompactifying the remaining
direction along a line of unbroken $SU(16)$ leads in the limit to $SO(32)$,
so these theories can also be regarded as compactifications of the $SO(32)$
string.

\subsection{Compactification to $D < 8$}

As $D$ decreases, more gauge groups become possible.  It is
convenient to take $\Gamma' = \Gamma^{SU(2)}$$\oplus$$ \Gamma^{9 -
D,9 - D}$. Here $\Gamma^{SU(2)}$ is the self-dual $SU(2)$ lattice
described above, $(p^9_3 ;\ \tilde p^9_3)$.  Let $w$, $\tilde w$
denote vectors in the weight lattice of $SO(18 - 2D)$.  Then
$\Gamma^{9 - D,9 - D}$ is defined as the lattice $(w ;\ \tilde w)$
such that
$w - \tilde w$ is in the root lattice of $SO(18- 2D)$. Momenta
$p_3$ are thus labeled $(w, p^9_3 ;\ \tilde w , \tilde p^9_3 )$.
Define an orthonormal set $w_{(i)}$ of $SO(18- 2D)$ vector
weights,
\begin{equation}
w_{(i)} \cdot w_{(j)} = \delta_{ij}.
\end{equation}
For $T$ take the shift\footnote
{For $D=8$ this is equivalent to the earlier construction under a
boost and Weyl reflection.}
\begin{equation}
v_2 = 0, \qquad v_3 = ( w_{(1)}, 0 ;\ 0, 0).
\end{equation}
Define the following:
\begin{eqnarray}
f_{(i)}:&& p_2 = u_{(i)}, \quad p_3 = (0,0 ;\ 0, \squ ),
\qquad i = 1, \ldots, 8, \nonumber\\
f_{(9)}:&& p_2 = 0, \quad p_3 = (0, \squ ;\ 0,0),
\nonumber\\
g_{(i)}:&& p_2 = u_{(i)},
\quad p_3 = (0,0 ;\ w_{(i)}, \squ ),\qquad i = 1, \ldots, 9 - D,
\nonumber\\
h_{(i)}:&& p_2 = 0, \quad p_3 = ( w_{(i)} , 0 ;\ 0, 0),
\qquad i = 1, \ldots, 9 - D.
\end{eqnarray}
All of these are allowed momenta, $g$ and $h$ being in the twisted
sector.  For $1 \leq n \leq 10 - D$ the set
\begin{equation}
f_{(i)},\ i = n, \ldots, 9, \quad
g_{(i)}, \ i = 1, \ldots n- 1, \quad
h_{(i)}, \ i = 1, \ldots 9 - D,
\end{equation}
is a basis of $18$$-$$D$ orthogonal vectors, with the $f$'s having
length-squared~2 and the $g$'s and $h$'s having length-squared~1.
By an $SO(18 - D,10 - D)$ transformation, bring the given set to
lie fully on the left.  The $f$'s form the long roots of
$Sp(20-2n)_1$ and the $g$'s and $h$'s form the short roots of
$SO(17 - 2D + 2n)_1$.  The remaining roots of each group (the sums
and differences of the short roots, and
$\half$ the sums and differences of the long roots) are all in the
untwisted sector. The product $Sp(20)_1$$\times$$SO(17 - 2D)_1$,
obtained first in the fermionic construction, is also in the same
moduli space.  Replacing $h_{(9 - D)}$ with
\begin{equation}
f_{(10)}:\quad p_2 = \sq (5,1^7), \quad
p_3 = (2 w_{(9-D)} , 0 ; 0, 3 \squ )
\end{equation}
gives the needed long root of $Sp(20)$, the additional short roots
being in the twisted sector.  Thus
we have found
\begin{equation}
Sp(20 - 2n)_1 \times SO(17 - 2D + 2n)_1
\qquad n = 0, \ldots, 10 - D
\label{sosp}
\end{equation}
at special points in a single moduli space.

The maximal symplectic group is $Sp(20)$.  The central charge of
$Sp(2n)_1$ is $(2n^2 + n)/(n+2)$.  The minimum central charge,
including an additional $18$$-$$n$$-D$ from $U(1)$'s needed to
saturate the left-moving rank, exceeds the available $26 - D$ if $n
> 10$. We also strongly believe that $SO(37 - 4D)$ is maximal.  It
might appear that one could get larger $SO(2n + 1)$ groups by
decompactification, say $SO(25)$ in
$D=4$ from $SO(25)$ in $D=3$.  However, this idea would lead to a
contradiction in going from $D=7$ to $D=8$, where we have already
shown that
$SO(5)$ is maximal.  In fact, study of these decompactifications
shows that decompactifying while preserving the larger orthogonal
symmetry gives a limit in which additional vectors become massless,
leading to a rank~16 Narain theory.  We expect that the $D=8$
result can be extended to show that $SO(37 - 4D)$ is maximal.

While larger $Sp$ and $SO$ groups cannot arise, there are other
possibilities. For example, twisting the $D$$=$$6$
$Sp(16)_1$$\times$$SO(9)_1$ theory by the additional
$T'$,
\begin{equation}
v' = \half(f_{(2)} + f_{(3)} + f_{(4)} + f_{(5)})
\end{equation}
leads to the gauge group $(F_4)_1 \times (F_4)_1 \times SO(9)_1$.
This is in the same moduli space: we could twist by
$T'$ before
$RT$, producing a different toroidal starting point which must
therefore be related by a boost~\cite{narain}. Similarly the group
$(F_4)_1 \times (F_4)_1 \times Sp(8)_1$ can be obtained~\cite{chl}.

\section{Discussion}

Twisting by the outer isomorphism reduces the rank by eight
relative to toroidal compactification, giving the moduli space of
the rank $-$$8$ theory. This orbifold construction appears to
reproduce all of the rank $-$$8$ solutions obtained in the
fermionic construction \cite{chl},\cite{sc} and shows that they
are  indeed special points within a single moduli space. A more
elegant characterization of the allowed lattices analogous to that
of ref.~\cite{narain} would be helpful.\footnote {The global {\it
reduction} of the rank (dimension) of a maximally supersymmetric
string moduli space via orbifolding should be contrasted with
familiar field theoretic mechanisms for symmetry enhancement at
specific points in a moduli space, such as Higgsing, whether by
fundamental scalars or more exotic composites.}

As mentioned in the introduction, this is only the first member in
a series of maximally supersymmetric theories characterized by
further reduction of the dimension of the moduli space, and it
remains to find a construction of the moduli spaces of the
remaining theories. Since Majorana fermion field theories only
contain $Z_2$ spin fields, it is possible that an asymmetric
orbifold construction of these moduli spaces will turn up
additional reduced rank moduli spaces that are absent in the
fermionic construction.  Conversely, it is also possible  that the
solutions with rank $r_L$$<$$10$$-$$D$ in four dimensions have no
orbifold realization. In these cases, the marginal deformations of
the six dimensional abelian torus are either absent or partly
constrained. These backgrounds do not appear to have a large radius
limit.

It is quite interesting to consider the implications of the CHL
string for weak/strong coupling duality. Dual theories describing
the strongly coupled limit of the heterotic string in various
dimensions have been proposed; see refs.~\cite{sdual},
\cite{ssdual},~\cite{eddual} and references therein.  In $D=4$ the
conjectured dual is the heterotic string itself~\cite{fontdual},
with the evidence being strongest in the case of toroidal
compactification~\cite{sdual},~\cite{eddual}.  What are the duals
of the CHL theories?  This presents a new challenge, because the
S-duality transformation includes electric-magnetic duality of the
low energy theory~\cite{montol}, and the groups $Sp(2n)$ and
$SO(2n+1)$ are not invariant under electric-magnetic duality but
rather are interchanged~\cite{gno}.  Thus, S-duality cannot leave
the individual points in the CHL moduli space invariant, but
requires that for each solution there be a heterotic string
solution with the dual group. The construction in the previous
section seems in no way to single out
$D=4$, and it is evident from the list~(\ref{sosp}) that it does not
automatically give dual groups.  But remarkably, just in $D=4$ where
required by S-duality, the series~(\ref{sosp}) is dual: the
strongly coupled behavior at one of these special points in moduli
space can be described by the weakly coupled theory with $n \to
6-n$, the point $n$$=$$3$ being self-dual. This is further evidence
for S-duality, apparently independent of previous results.
Moreover it is evidence for duality of the full string theories,
not the low energy effective theories, because stringy phenomena
(enhanced gauge symmetries) play an essential role. It would be of
course interesting to compare the spectrum of short multiplets of
these two theories.

In $D=6$ there is a puzzle, however.  There is mounting evidence
for string-string duality, the heterotic string on $T^4$ being
equivalent to the IIA string on $K_3$
\cite{ssdual},~\cite{eddual}.  It is natural to look for a related
dual in the present case. Indeed, Ferrara, Harvey, Strominger, and
Vafa~\cite{andy} have recently considered a similar situation, in
which
 the IIA string is compactified on $K_3$$\times$$T^2$ modded by a
$Z_2$ isomorphism.  They argue that the dual theory is the
heterotic string on an asymmetric orbifold of $T^6$, where the
reflection includes an outer isomorphism of $E_8$$\times$$E_8$.  In
that case the reflection also acts on the right-movers, breaking
half the supersymmetry, but it suggests that the dual in our case
as well should be some twisting of the $K_3$ compactification of
the IIA string.\footnote{We would like to thank Paul Aspinwall and
Andy Strominger for independently suggesting this and pointing out
the parallel with ref.~\cite{andy}.}  The following argument of
Seiberg~\cite{seiberg} appears to exclude this attractive
possibility.  In order to obtain $N$$=$$2$, $D$$=$$6$ supersymmetry
it is necessary  to preserve the $(4,4)$ world-sheet supersymmetry
of the $K_3$ theory~\cite{bd}.  This is also a good background for
the IIB string~\cite{seiberg}, giving chiral
$N$$=$$2$ $D$$=$$6$ supergravity.  Spacetime anomaly cancellation
then determines completely the massless spectrum, in particular the
number of massless vectors and moduli in the IIB theory.  For the
IIA on the same background this implies a rank~24 gauge group,
exactly as found in toroidal compacitication of the heterotic
string but inconsistent with theories of reduced rank.  But there
are other possibilities; for example, the dual could be a type~II
theory with both spacetime supersymmetries right-moving,
corresponding to a right-moving torus times a generic $N$$=$$1$
left-moving theory.  The full picture of duality of the CHL
theories remains to be discovered.

\subsection*{Acknowledgments}

We would like to thank Paul Aspinwall, Jeff Harvey, Costas Kounnas,
Joe Lykken, and Andy Strominger for illuminating discussions. S.C.
acknowledges the warm hospitality of the CERN Theory division during
the completion of this work. This work is supported by NSF grants
PHY91-16964 and PHY94-07194.

\subsection*{Appendix}
\setcounter{equation}{0}
\def\theequation{A.\arabic{equation}}

For the lattice used in $D=8$ we show that if
\begin{eqnarray}
&& p_1 = p_1' = 0 \label{p1van}\\
&& p \cdot p = 2 \bmod 4,\quad
p' \cdot p' = 2 \bmod 4, \label{mod4}
\end{eqnarray}
then
\begin{equation}
p \cdot p' = 0 \bmod 2. \label{dot}
\end{equation}
Eq.~(\ref{p1van}) implies that $p_2$ and $p_2'$ lie on $\sqrt{2}
\Gamma^8$ and so contribute to the lengths~(\ref{mod4}) only modulo
4 and to the dot product~(\ref{dot}) only modulo 2.  Similarly the
$RT$ projection implies that $n_9 + \tilde n_9$ is a multiple of~4,
hence $p^9$ contributes in the same way.  Finally, $n_8 = 2r-\tilde
n_8$ for integer $r$.  Then $p \cdot p \bmod 4$ is $\half
(n_8^2 - \tilde n_8^2) = 2r(r - \tilde n_8)$.  It follows
from~(\ref{mod4}) that $r$ is odd and
$n_8$ and $\tilde n_8$ even.  Similarly $n'_8$ and $\tilde n'_8$
are even, and so eq.~(\ref{dot}) holds.

For long roots $p_1 = 0$ and $p \cdot p =2$ with $p$ purely
left-moving and so having a positive inner product.  It follows from
this and eq.~(\ref{dot}) that distinct long roots are orthogonal.

\end{document}